\documentclass[sigconf]{acmart}
\AtBeginDocument{%
  }

\copyrightyear{2026}
\acmYear{2026}
\setcopyright{cc}
\setcctype{by}
\acmConference[ICSE-FoSE]{2026 IEEE/ACM 48th International Conference on Software Engineering: Future of Software Engineering }{April 12--18, 2026}{Rio de Janeiro, Brazil}
\acmBooktitle{2026 IEEE/ACM 48th International Conference on Software Engineering: Future of Software Engineering (ICSE-FoSE), April 12--18, 2026, Rio de Janeiro, Brazil}
\acmPrice{}
\acmDOI{10.1145/3793657.3793903}
\acmISBN{979-8-4007-2479-4/2026/04}

\begin{document}

\title{Political and Ideological Pressure in Software Engineering Research: The Case of DEI Backlash}

\author{Sonja M. Hyrynsalmi}
\email{sonja.hyrynsalmi@lut.fi}
\orcid{0000-0002-1715-6250}
\affiliation{%
  \institution{LUT University}
  \city{Lahti}
  \country{Finland}
}
\author{Chris Brown}
\email{dcbrown@vt.edu}
\orcid{0000-0002-6036-4733}
\affiliation{%
  \institution{Virginia Tech}
    \city{Blacksburg}
  \country{United States}}

\author{Alexander Serebrenik}
\email{a.serebrenik@tue.nl}
\orcid{0000-0002-1418-0095}
\affiliation{%
  \institution{Eindhoven University of Technology}
    \city{Eindhoven}
  \country{The Netherlands}}

\author{Sebastian Baltes}
\email{sebastian.baltes@uni-heidelberg.de}
\orcid{0000-0002-2442-7522}
\affiliation{%
  \institution{Heidelberg University}
    \city{Heidelberg}
  \country{Germany}}

  \author{Letizia Jaccheri}
\email{letizia.jaccheri@ntnu.no}
\orcid{0000-0002-5547-2270}
\affiliation{%
  \institution{Norwegian University of Science and Technology}
    \city{Trondheim}
  \country{Norway}}



\begin{abstract}
Political and ideological pressures shape global research. 
Recently, these pressures have become particularly visible in research related to diversity, equity, and inclusion (DEI).
Drastic changes in national funding and governmental guidance, especially in the US, have affected the global software engineering research ecosystem. 
The impacts of these pressures on research are not always direct, as they operate at multiple levels. 
However, what is clear is that these pressures affect every field, including software engineering (SE), despite the belief that our field is politically and ideologically neutral. In this position paper, we examine cases of political and ideological pressures on the SE research ecosystem. We investigate the community's perceptions of political and ideological pressures by analyzing community survey responses and outlining case examples of DEI backlash in SE research across three levels: macro, meso, and micro. 
Our research shows how recent political and ideological pressures have affected SE research across these levels, and, as a result, we propose actionable steps for the community to address these issues at different levels.
\end{abstract}


\maketitle

\section{Introduction}
Software engineering (SE) is not a politically or ideologically neutral field. 
Historically, a large amount of funding, motivation, and institutional support for SE research emerged from Cold War interests. For example, the term `software engineering', although previously used by Margaret Hamilton and Anthony Oettinger, was globally established with the 1968 NATO conference~\cite{edwards1996closed, kaplan2016dark}.

Political and ideological pressures can take many forms and focus on different issues. Right now, the biggest political and ideological pressure comes from changes in US politics. For instance, one of the main areas targeted by the current administration is diversity, equity, and inclusion (DEI). On January 21, 2025, an Executive Order titled ``Ending Illegal Discrimination and Restoring Merit-Based Opportunity'' (the ``January 21 DEI Order'') was issued. Before that, on January 20, two DEI-related executive orders titled ``Defending Women From Gender Ideology Extremism and Restoring Biological Truth to the Federal Government'' (the ``Gender Order'') and ``Ending Radical and Wasteful Government DEI Programs and Preferencing'' (the ''January 20 DEI Order'') were enacted~\cite{High2025, WhiteHouse2025a, WhiteHouse2025b}. 

Diversity is a critical part of SE, enhancing software development and innovation~\cite{shonan}. 
Previously, global software companies such as Amazon, Meta (Facebook), Alphabet (Google), and Accenture were forerunners in DEI actions, but have now reduced or renamed DEI initiatives~\cite{MurrayBohannon2025, AllenFischer2025, AllenFischer2025b, Zuckerberg2025}. This situation, widely described as a ``DEI backlash''~\cite{thompson2024coping}, is not new. However, this is the first time that DEI initiatives have faced backlash due to changes to laws previously enacted to advance DEI\cite{hyrynsalmi2025tech}. Recent research shows that software companies try to justify changes in their DEI initiatives with `business' reasons, but there are also strong political reasons behind~\cite{hyrynsalmi2025tech, desantos2025diverse}. This DEI backlash has also affected scientific research. For example, the National Science Foundation (NSF) disrupted approximately 2,000 grants and \$697 million in research funding\footnote{\url{https://grant-witness.us/nsf-data.html}}---including programs related to broadening participation in computing. Nevertheless, some software companies and researchers choose to make it clear that they support DEI initiatives, even as they face strong political and ideological pressure~\cite{hyrynsalmi2025tech}.

\section{Scope and Empirical Grounding}

This position paper aims to raise awareness of political and ideological influences on SE research. We discuss recent examples from SE research in which political and ideological pressures, particularly those related to DEI initiatives, affect the research community.
We chose to focus on DEI-related cases, as there is a strong ideological and political movement against DEI right now. Although this current movement is US-based, our examples will illustrate that it has a global impact on SE research.

As background data, we use survey results collected for the Future of Software Engineering track at ICSE 2026.\footnote{https://zenodo.org/records/18217799} The survey was distributed to SE researchers from October to November 2025, and yielded 280 completed responses. Through thematic analysis~\cite{braun2006using}, we identified and categorized responses that highlight political and ideological aspects within the SE research community.

\section{Survey Insights from the SE Community}

In this section, we discuss responses in which respondents explicitly referenced political or ideological challenges within the SE research community. The survey itself did not include direct questions about political or ideological pressure. However, respondents raised these issues in their responses to several open questions. For that reason, we analyzed the entire dataset and identified four questions in which respondents exhibited a political or ideological tone. We use the terms ``political'' and ``ideological'' to describe situations in which researchers or research practices are influenced by aspects such as policy agendas or political theories (e.g., liberalism, conservatism, or nationalism), rather than by disciplinary standards or peer review.

\textbf{Question `What aspect or aspects of the software engineering research community do not work well, and why?''} One respondent explicitly mentioned political and ideological agendas in conferences:

\begin{quote}
    ``some conferences pushing a political (usually woke) agenda'' (192723543)
\end{quote}

The respondent did not clarify what kind of pressure exists, however the use of the term `woke' implies they would see DEI initiatives in SE conferences potentially in a negative way.

\textbf{Question ``If you could make one change, what would you change and what outcome from that change would you like to see in the software engineering research community?''} Two responses proposed changes to mitigate political or ideological pressures in the SE research community. Interestingly, both of these responses strongly focused on a geographical approach. 

\begin{quote}
    ``avoid dependency of American editorials'' (196281159)
\end{quote}

\begin{quote}
    ``Lobby more at EU level'' (196290770)
\end{quote}

Unfortunately, nothing more was written in these responses, so we do not know specifically what the respondents wanted to be lobbied at the EU level (e.g., funding) or the reasons for avoiding dependency on American editorials (e.g., costs, politics, etc.).  

\textbf{Question ``What aspects of being in the software engineering research community cause the greatest amount of stress for you?''} Two participants mentioned aspects of political and ideological pressures invoke stress within the SE research community. One respondent highlighted cultural and political differences:  

\begin{quote}
    ``The politics and cultural differences that cause friction. The barriers to acceptance for different methods, different ways of thinking, different ideas and approaches.'' (196119370)
\end{quote}

Another participant noted that challenges associated with the degradation of research in politics, including obtaining funding and international collaborations, are the greatest stressors.
\begin{quote}
    ``Political antagonism to research and the associated problems with funding and international collaboration'' (194327041)
\end{quote}

There were also several mentions of conference costs being too high and remaining a barrier for researchers all over the world to attend. One participant called that 'economic apartheid':
\begin{quote}
    ``The increasing number of papers to review (service in research conferences and journals). Pressure to publish Anxiety epidemic among students Expenses to participate in international conferences. The scientific world is a real economic apartheid.'' (196352856)
\end{quote}

\textbf{Question ``When it comes to your current or future mentees (e.g., colleagues, junior faculty, postdoctoral researchers, students, fellow students), what concerns you the most about their future as members of our community and your ability to prepare them for it?''}. Multiple respondents expressed concerns related to political and ideological pressures regarding their mentees' futures in the SE research community. One participant explicitly mentioned concerns about:

\begin{quote}
    ``Attacks on diversity and inclusion in the US'' (193668599)
\end{quote}

This was repeated in another response, highlighting both diversity and inclusion issues related to younger researchers' identity and visa status:  
\begin{quote}
    ``Then there is the question of visa, sometimes my students cannot get a visa for a location. I also don't think we should be hosting any conferences in places that are hostile to foreigners, women, or LGBTQIA+ (e.g., the USA). I don't want to attend anywhere where my human rights are up for debate, or where my colleagues have that experience'' (192349636)
\end{quote}

Another respondent discussed the overall anti-science movement and its potential to inhibit rising researchers from obtaining funding or making an impact:
\begin{quote}
    ``In addition, the widespread loss of public respect for science and associated resistance to funding it and following it'' (194327041)
\end{quote}

Overall, these responses demonstrate the existence of political and ideological backlash within the SE research community. Next, we present three cases to illustrate how the backlash manifests across different levels and offer suggestions for counteractions.  

\section{Case Examples}
To supplement the survey findings, we provide example cases of DEI backlash to demonstrate the influence of political and ideological pressure in our research field~\cite{blalock1960social}. For this position paper, we identified three cases that faced political and ideological pressure, were from the software engineering discipline, and focused on recent experiences within the past year. We discuss these three examples at the micro, meso, and macro levels to illustrate the complexity of political and ideological pressures in SE research. At each level, we illustrate how these pressures have affected software engineering research or researchers and offer suggestions for how the SE research community can pursue an alternative future under such pressures. 

\subsection{Micro-level Case: Editor Requests Removal of DEI Terminology}

The micro-level refers to outcomes, interactions, or experiences at the individual level. In the context of SE research, this includes paper submission, review, and editorial decision-making processes within journals and conferences. 

\subsubsection{Case}

An example of micro-level political and ideological pressure concerns a recent paper about DEI aspects in software engineering.
The paper, authored by European researchers, was submitted to a top-tier software engineering journal published by a US-based organization consisting of a global network of technology professionals.
\footnote{\url{https://www.linkedin.com/posts/sonjahyrynsalmi\_recently-new-us-regulations-have-started-activity-7360726003104591875-8jgp}}
However, the journal’s decision was influenced by political pressure. Specifically, recently issued executive DEI orders led editors to hesitate about whether such a clearly DEI-related paper could be published without modifications. 

The executive order pertained to US government funding of DEI work~\cite{WhiteHouse2025a}, and had no scope over what anyone can publish.
However, the editor of a journal decided to approach researchers who had submitted to the journal. He explained that the editorial board was supportive of the paper continuing in the process, but requested changes to minimize DEI content. They motivated the change request with a message: ``\textit{As you may well be aware, here in the United States our present government has entered into a campaign to stop diversity, equity, and inclusion (DEI) programs and initiatives in all areas where the US government has involvement. The following DEI Executive Order (EO) was issued by the Trump administration: Ending Radical and Wasteful Government DEI Programs and Preferencing.}'' They also added: ``\textit{-The publisher- has stated that they will continue to support Editors-in-Chief and editorial boards on publication decisions, but as a tax-exempt organization operating under US Code, we are not immune from this executive order. We have to be wary of putting our editors at risk if we publish a paper that could be portrayed as being in non-compliance}''. \footnote{Specific identifying details have been intentionally omitted to protect individuals involved}

The editor also implicated which parts in the text were troubling, such as: ``\textit{P8 C1 L47: drop `does not encourage diversity in higher positions in academia' for DEI EO reasons. P8 C1 L57: more DEI fixes...not `ensuring diversity' —> `ensuring accuracy'. Also P8 C2 L21 `promoting unbiased representation across' P8 C2 L24: whole paragraph serves up DEI issues, need to soften. Mentioning `gender identity' is another hot button target of the present US administration, and other topics like `abelism' and `fatphobia' that are concerns they characterize as `woke'}''.

The authors respectfully declined to make changes that would have removed core aspects of their findings and analysis. The paper was later accepted for publication at another software engineering venue outside the United States. Importantly, this case is not presented as a critique of individual editors or the journal, who acted transparently and in good faith under legal and employment-related risks and pressure. Instead, it highlights how macro-level political decisions can translate into micro-level actions that affect the SE research community---in particular editorial practices, authorial freedom, job safety, and, overall, the perception that some topics are riskier to publish than others. 

\subsubsection{Suggestions for the SE Community}\ \\
{
We suggest that SE journals \textbf{clearly communicate} that they consider \textbf{research on politically and ideologically sensitive topics}, such as DEI, to be \textbf{within scope}.
Editorial boards and publishers could develop mechanisms to \textbf{support editors} who are afraid to lose their employment or their position on the editorial board when handling politically or ideologically sensitive submissions. 
Publishers and associations could offer \textbf{training for reviewing of and publishing} about politically and ideologically sensitive topics. For example, training on how to navigate publication, review, and communication practices in politically sensitive contexts.
Associations could \textbf{openly discuss whether there is a risk of self-censorship}, i.e., that researchers would start to avoid certain topics or terminology altogether. There could be specific events, workshops, tracks or special issues to support researchers in researching `risky' topics.

%
%
%

\subsection{Meso-Level Case: Conference Relocation}

Meso-level refers to outcomes, interactions, or connections at the community or organization level. In SE research contexts, this can refer to conferences, departments, or associations. 

\subsubsection{Case}
For this level, we examine the case of Code4Lib\footnote{\url{https://code4lib.org/}}, an organization of programmers and technologists who support the development of open technologies for libraries, museums, and archives. The organization holds an annual conference for and by computer programmers and library technologists. However, Code4Lib 2026 was forced to relocate to a new venue due to a diversity scholarship.\footnote{\url{https://lists.clir.org/cgi-bin/wa?A2=ind2512&L=CODE4LIB&P=19235}}  Their first post about the situation described challenges with DEI executive orders and how the host university, Carnegie Mellon University, reacted to those: \textit{``Recently, the U.S. Department of Education Office for Civil Rights (OCR) ordered Carnegie Mellon University to review all third-party partnerships for compliance with Title VI of the Civil Rights Act. As part of this review, the university determined that aspects of Code4Lib’s diversity scholarship program raised compliance concerns under Title VI and Title IX. In light of these findings, CMU has concluded that it cannot move forward as a host site. The LPC and Scholarship Committees explored whether adjusting the location or presentation of scholarship information might address these issues, including shifting content to external domains. However, it was determined that these changes would not resolve the underlying concerns related to compliance with federal, state, and local laws, as well as university policy. As a result, the university is unable to host, sponsor, or partner with a conference whose scholarship structure does not meet these legal requirements.''}

In later emails, the organizing committee members highlighted that they are not making any changes to the scholarship program, and as a result, the conference was required to change venue. This case shows how macro-level laws and actions act in meso-level decisions, such as where to host a conference. Beyond Code4Lib, many other tech-related conferences (e.g., Collision Conference~\cite{collision}, InCyber Forum~\cite{incyber}, and Grace Hopper~\cite{gracehopper}) and scientific meetings  (e.g., the  International Association of Cognitive Behavioral Therapy (IACBT), International Society for Research on Aggression (ISRA), and American Society for Clinical Pathology (ASCP)~\cite{skift}) have been canceled or relocated due to US executive orders. 

\subsubsection{Suggestions for the SE Community} \ \\  
Overall, if DEI initiatives cannot be executed or if there is caution about how openly organizations can support them, we should, as a community, \textbf{ask whether it is worthwhile to hold conferences in the US} right now. Furthermore, the SE community, especially conference organizing and steering committees, need to have an open discussion about conference venues and how to promote more accessible conferences to avoid political and ideological pressures (e.g., visa issues, discriminatory practices, etc.) that prevent the community from getting together. Conference costs, including travel and accommodation, are also a barrier to bringing the community together. 

Code4Lib's organizing committee communicated openly with their community about their values and did not compromise them. They decided to make a (likely costly) change in location rather than abandon their DEI initiatives. \textbf{Communicating transparently about values and their support} is something we can do at different levels in our community. 

%

\subsection{Macro-Level Case: NSF Funding Cuts}

The macro level refers to global or national outcomes or interactions. In SE research contexts, this may be national funding, national laws that guide universities, or global agreements and laws. 

\subsubsection{Case}
For the macro-level case, we discuss recent funding cuts and changing priorities at the National Science Foundation. NSF is the primary funder for research in the US, spending approximately \$1 billion dollars on computing research in 2024~\cite{nsf}. However, recent changes enacted by the current administration, including proposed budget cuts, staff restructuring, and canceling grants related to DEI, have had significant impacts on research in general and SE research in particular. One example is the recent de-funding of Gendermag.
The following is based on an interview with Margaret Burnett~\cite{burnett-interview-2026}, Co-Director of the GenderMag Project.\footnote{\url{https://gendermag.org/}}

GenderMag (Gender Inclusiveness Magnifier)~\cite{burnett2016gendermag} is a method for software practitioners to use to find and remove ``inclusivity bugs'' from the software they are developing.
From a requirements engineering perspective, it relates to the non-functional requirement of usability for widely diverse users~\cite{guizani2020requirements}.
From a software process perspective, it is a software process that enables early-stage, systematic evaluation of early prototypes, and also serves as a debugging methodology~\cite{guizani2022debug}.
From an SE business perspective, it aims to increase market share.
Its results have been empirically shown to be very effective (e.g.,~\cite{murphyhill2024gendermag, vorvoreanu2019gender, anderson2026findfix, guizani2022debug, hamid2025-tiis}).

In 2025, shortly after the new administration took office, two NSF grants funding the GenderMag project were abruptly rescinded.  
This happened even though the annual project reports, showing productive new results, had been approved every year (the most recent approval just a few weeks before the grant was rescinded). 

From a micro-level, this left the GenderMag project unfunded. 
But the national, meso-level impacts were much larger. GenderMag was not the only DEI-related research project that was defunded.
More than 1,500 grants totaling over \$1 billion, most related to DEI, have also been defunded~\cite{urban}, leaving researchers across the U.S. suddenly without funds to support their work.
In addition, replacement funding prospects suddenly disappeared.  
The NSF changed their ``priorities'' to remove anything DEI-oriented.\footnote{\url{https://www.nsf.gov/updates-on-priorities}}
Further, the NSF's funding dropped by 25 percent~\cite{newYorkTimes-2025}, making funding much harder to secure.
Turning to private foundations, which normally fund inclusion-related work, offered little relief, because the sudden influx of now-defunded researchers overwhelmed their capacities. For example, the Spencer Foundation saw a five-fold increase in proposals~\cite{spencerFoundation-2025}.

Ultimately, some researchers had to change research directions in order to continue to support their students and publish.
Evidence of this is the switch from a long-term upward trend in SE publications related to DEI to a sudden decline.
According to Google Scholar, publication rates dropped from ``about 36,700'' papers in 2024 to``about 33,000'' papers in 2025, an estimated 10\% decline in the span of just one year.
This is in stark contrast to the prior year's \textit{increase} of about 2\%, the increase the year before that of about 12\%, and the increase the year before that of about 6\%.%
\footnote{As per Google Scholar advanced searches on Jan. 3, 2026: With the exact phrase: ``software engineering''; with at least one of the words: ``inclusion equity diversity DEI''; anywhere in the article; separately for year ``2025-2025'', ``2024-2024'', and so on.} 

\subsubsection{Suggestions for the SE community}\ \\
We need to find, communicate, and lobby for \textbf{more funding mechanisms to connect researchers across continents}. Global SE associations could design international funding mechanisms that can temporarily support researchers affected by abrupt, politically driven national funding withdrawals.

We need \textbf{explicit reviewing guidelines} for SE reviewers to every conference, funding and journal site, disallowing political ideologies as review criteria---and PC chairs and editors enforcing this.
For example, journals and conferences could include a statement like this one in ACM's Peer Review Policy\footnote{\url{https://www.acm.org/publications/policies/peer-review}}: ``ACM requires that the peer review process and related decisions be free of bias based on nationality, religious or political beliefs, gender or other demographic characteristics, personal or professional conflicts or competing interests''.

%
%

\section{Discussion}

The signals in the survey results and case examples indicate that political and ideological pressures are impacting the SE research community. These influences affect different levels of the community and thus require counteractions at different levels. Recent studies reveal that some key players in the tech industry have decreased their support for DEI initiatives for political and ideological reasons~\cite{hyrynsalmi2025tech, desantos2025diverse}. However, our responsibility as researchers is to explore SE research topics that may have a societal or political impact.

Identifying and talking about political and ideological pressure is not always easy. However, it can directly affect the research directions of the field. Fighting against this kind of pressure requires clear and concrete actions, such as peer-review policies or financial support for underrepresented groups. Our suggestions focus on how, for example, publishers, associations, organizing committees, and individuals can play their part in this issue. Still, the most powerful tool is communication: speaking about our values, not renaming actions or trying to hide them, and maintaining open communication shows that the community supports researchers facing political and ideological pressures. Furthermore, we need to be better prepared as a community when conducting research on sensitive topics, as the findings of such studies could face strong opposition for political or ideological reasons. 

Although we focus on DEI backlash, the mechanisms identified are topic-agnostic. Any area of computing and SE research can become vulnerable once it intersects with political or ideological interests. For example, the US administration has also targeted research on combating misinformation and disinformation online.\footnote{\url{https://www.nsf.gov/updates-on-priorities\#misinformation}} For that reason, the SE community needs explicit safeguards, such as reviewing guidelines or targeted funding, to ensure that we continue to produce impactful research---especially in politically and ideologically challenging times. 



\begin{acks}
We thank Margaret Burnett for generously sharing her time and insights for this position paper and Ronnie de Souza Santos for his valuable support and constructive comments.
\end{acks}

\bibliographystyle{ACM-Reference-Format}
\bibliography{software}
\end{document}